\UseRawInputEncoding
\documentclass[pra,aps,twocolumn,epsfig,superscriptaddress,showpacs,showkeys]{revtex4}
\usepackage{amsmath}
\usepackage{amssymb}
\usepackage{graphicx}
\usepackage{dcolumn}
\usepackage{bm}
\usepackage[colorlinks=false,dvipdfm]{hyperref} 
\usepackage{setspace}
\usepackage{amsfonts}
\usepackage{rotating}
\usepackage{makeidx}
\usepackage{amsthm}
\usepackage{amsmath}
\usepackage{color}

\begin{document}
\title{Discrimination of Coherent States via Atom-Field Interaction without Rotation Wave Approximation\footnote{Commun. Theor. Phys. 75 (2023) 065104}}

\author{Jin-Hua Zhang}
\affiliation{Department of
Physics, Xinzhou Teachers University, Xinzhou 034000, China}

\author{Fu-Lin Zhang}
\email[Corresponding author: ]{flzhang@tju.edu.cn}
\affiliation{Department of Physics, School of Science, Tianjin
University, Tianjin 300072, China}

\author{Mai-Lin Liang}
\affiliation{Department of Physics, School of Science, Tianjin
University, Tianjin 300072, China}

\author{Zhi-Xi Wang}
\affiliation{School of Mathematical Sciences, Capital Normal
University, Beijing 100048, China}

\author{Shao-Ming Fei}
\email[Corresponding author: ]{feishm@cnu.edu.cn}
\affiliation{School of Mathematical Sciences, Capital Normal
University, Beijing 100048, China}
\affiliation{Max-Planck-Institute for Mathematics in the Sciences,
D-04103 Leipzig, Germany}


\begin{abstract}
The quantum state discrimination is an important part of quantum information processing. We investigate the discrimination of coherent states through Jaynes-Cummings (JC) model interaction between the field and the ancilla without rotation wave approximation (RWA). We show that the minimum failure probability can be reduced as RWA is eliminated from JC model and the non-RWA terms accompanied by the quantum effects of fields (e.g. the virtual photon process in the JC model without RWA) can enhance the state discrimination. The JC model without RWA for unambiguous state discrimination is superior to ambiguous state discrimination, particularly when the number of sequential measurements increases. Unambiguous state discrimination implemented via the non-RWA JC model is beneficial to saving resource cost.
\end{abstract}




\maketitle





\section{Introduction} \label{intro}
Quantum state discrimination plays important roles in quantum information processing \cite{Enk2002PRA}. Since it is impossible to discriminate two non-orthogonal quantum states perfectly, it is then a key task in state discrimination to find out the maximal success probability. The state discrimination problems can be divided into two categories:
ambiguous \cite{Helstrom1976,Wittmann2008,Tsujino2011,Assalini2011,Xiong2018JPA,Xiong2019PRA} and unambiguous quantum state discrimination
\cite{Ivanovic1987PLA,Peres1988PLA,Dies1988PLA,Bennett1992PRL,Bergou2003PRL,Pang2009PRA,Pang2013PRA,Bergou2013PRL,Namkung2017PRA,Namkung2018SR,Zhang2017arXiv}.
The lower bound of ambiguous state discrimination is given by Helstrom primarily in \cite{Helstrom1976}. This bound can be saturated theoretically \cite{Wittmann2008,Tsujino2011,Assalini2011}.
The error-free unambiguous state discrimination plays key roles in various contexts in quantum information theory, which can also be optimized through different channels \cite{Pang2013PRA,Bergou2013PRL,Namkung2017PRA,Namkung2018SR,Zhang2017arXiv,JHZhang2020PRA,JHZhang2020Entropy}.

Especially, as a kind of physical implementation of state discrimination, the discrimination of nonorthogonal coherent-state signals has attracted much attention \cite{Silberhorn2002PRL,Lorenz2004APB,Lance2005PRL,Takeoka2005PRA,Takeoka2008PRA,Bergou2010JMO,Wittmann2010PRL,Wittmann2010PRA,
Weedbrook2012RMP,Becerra2013Nap,Becerra2014Nap,Sych2016PRA,Wittmann2010JMO,Tsujino2011PRL,M¨¹ller2015NJP}. Since the coherent states are easy to generate and have the best achievable signal-to-noise
ratio during the propagation of information \cite{Han2018NJP}, the discrimination of coherent states is of great importance in quantum optics. Nevertheless, concerning the existing experiment protocols in physical implementation, there remains a significant
gap between the ideal bound and the possible optimal result of the experiment for both ambiguous and unambiguous discriminations.

In order to reduce the deficiency, the coherent light states are coupled with an auxiliary atom and discriminated by measuring the ancilla. This is equivalent to performing a positive operator-valued measure (POVM) on the light states according to the Naimark theorem. The interaction between the field and the atom can be characterized via the Jaynes-Cummings (JC) model \cite{Han2018NJP}. The general JC model is difficult to be solved accurately. Under rotation wave approximation (RWA), where the non-RWA terms in the Hamiltonian are neglected, an exact solution of the JC model can be obtained. The light state discrimination implemented via the JC model under RWA is discussed in \cite{Han2018NJP}. If the post-measured light state is not thoroughly destroyed, the left information can be used for another measurement.
These sequential measurements are performed on a sequence of auxiliary atoms. This nondestructive scheme with sequential measurement has been successfully applied in \cite{Han2018NJP} to reduce the failure probability and acquire results approaching the ideal bound.


In order to get more accurate and practical results, the non-RWA terms in the atom-field Hamiltonian should be reserved. Namely, we consider the JC model without RWA. This non-RWA JC model can be solved by perturbation method  \cite{Peng1992PRA,Peng1993PRA}.
It also includes the virtual photon field causing the lamb shift and quantum
fluctuations \cite{Peng1992PRA,Peng1993PRA}.
it is of great importance in quantum optics, e.g. in ensuring the causality of atom-field coupling system. 

From a theoretical point of view, JC model under RWA is the simplest model that describes the interaction between atom and light field. Thus, the calculation for state discrimination of RWA gives only a rough estimation for the protocol
based on JC model. The RWA can be employed if the coupling between the atom and the light field is quite weak. From an experimental viewpoint, the JC model can be simulated by Josephson
charge qubit coupling to an electromagnetic resonator \cite{Wallraff2004Nature, Simmonds2004PRL}, a superconducting quantum interference device coupled with
a nanomechanical resonator \cite{Chiorescu2004Nature,Johansson2006PRL} and an
LC resonator magnetically coupled to a superconducting qubit
\cite{Forn2010PRL}. With the progress of experimental technology, these artificial atoms may interact with on-chip resonant circuits \cite{Wallraff2004Nature,Simmonds2004PRL,Chiorescu2004Nature,Johansson2006PRL,Yu2003Science,Forn2010PRL} very strongly, and the RWA cannot describe well the strong-coupling regime \cite{Liu2009Eru}.  Hence, from the above two aspects, it is also very desirable to investigate state discrimination based on the model without RWA.

The present study aims to find out whether light states can be discriminated with a better result in a strong-coupling regime. Namely, we tend to optimize the coherent state discrimination to a further step and find a minimum failure probability which is more closer to the ideal lower bound in the framework of non-RWA JC model than the existing results with RWA in \cite{Han2018NJP}.  In addition, we also intend to  find out the roles played by QEOF, which are included in the JC model without RWA, in quantum state discrimination.

The paper is organized as follows. First, we present an optimal ambiguous discrimination of two coherent states via JC model without
RWA. We show that the minimum probability of failure is reduced and the ideal Helstrom bound can be approached further compared with the results for the JC model with RWA. In addition,  we study a protocol for the physical implementation of unambiguous discrimination of coherent states including non-destructive sequential measurements. We show
that under the effect of non-RWA term, the discrimination of light states can be enhanced.
 We illustrate the relations between such superiority and the number of sequential measurements. We summarize in the last section.

\section{Ambiguous discrimination}\label{Ambiguous discrimination}

We first consider the ambiguous discrimination of the well-known coherent states $|\psi_1\rangle=|\alpha\rangle$ and $|\psi_2\rangle=|-\alpha\rangle$ occurring with probabilities $P_1$ and $P_2$, respectively, where
\begin{equation}
|\pm\alpha\rangle=\sum\limits_{n=0}^{\infty}F_n(\pm\alpha)|n\rangle,
\end{equation}
with $F_n(\pm\alpha)={\rm{e}}^{-|\alpha|^2/2}\frac{(\pm\alpha)^n}{\sqrt{n!}}.$
In order to discriminate  $|\psi_1\rangle$ from $|\psi_2\rangle$ ambiguously, we couple the system with
an ancilla state $|k\rangle$ and perform a joint unitary transformation $U$ \cite{Namkung2018SR},
\begin{eqnarray}\label{unitary operator for ambiguous discrimination}
U|\psi_1\rangle|k\rangle=\sqrt{1-r_1}|\chi_1\rangle|1\rangle+\sqrt{r_1}|\phi_1\rangle|2\rangle, \nonumber\\
U|\psi_2\rangle|k\rangle=\sqrt{r_2}|\chi_2\rangle|1\rangle+\sqrt{1-r_2}|\phi_2\rangle|2\rangle,
\end{eqnarray}
where $\{|1\rangle,|2\rangle\}$ is an orthogonal basis of the ancilla. Through a von Neumann measurement $\{|1\rangle\langle1|,|2\rangle\langle2|\}$ on the ancilla, the states $|\psi_i\rangle$ are identified with error probability
$$P_{\rm{err}}=P_1r_1+P_2r_2,$$
where $r_1$ and $r_2$ satisfy $\langle\psi_1|\psi_2\rangle=\sqrt{(1-r_1)r_2}\langle\chi_1|\chi_2\rangle+\sqrt{(1-r_2)r_1}\langle\phi_1|\phi_2\rangle$ according to Eq. (\ref{unitary operator for ambiguous discrimination}). By a straightforward calculation one can see that the Helstrorm bound (minimum failure probability) $P_A=\frac{1}{2}(1-\sqrt{1-4P_1P_2 s^2})$ with $s=|\langle\psi_1|\psi_2\rangle|$ is saturated when $|\chi_1\rangle=|\chi_2\rangle$ and $|\phi_1\rangle=|\phi_2\rangle.$

The detailed physical system contains light fields (coherent states) interacting with
a two-level atom prepared in the ground state $|g\rangle$ which plays as the auxiliary qubit in state discrimination.

In order to get more accurate and practical results and find out the roles
played by QEOF in state discrimination,  we consider the JC model without RWA.
Namely, we do not neglect the counterrotating terms in the Hamiltonian \cite{Peng1992PRA},
which can be written as
\begin{equation}
H=H_0+H_{I_0},
\end{equation}
with
\begin{equation}
H_0=\omega a^{\dagger}a+\frac{1}{2}\hbar\omega_0\sigma_z,
\end{equation}
and
\begin{equation}\label{Effective Hamiltonian1}
H_{I_0}=g(\sigma_{+}a+a^{\dagger}\sigma_{-}+a^{\dagger}\sigma_{+}+a\sigma_{-}),
\end{equation}
where $\omega$ is the frequency of the light field, $\omega_0$ is the frequency of the atomic transition, $\sigma_+$ and $\sigma_-$ denote the
atomic raising and lowering operators, $a$ and $a^{\dagger}$
are the field annihilation and creation operators, $g$ is the atom-field coupling constant. We focus on the situation where the dipole coupling is on resonance, i.e., $\omega= \omega_0$. The last two terms in Eq. (\ref{Effective Hamiltonian1}) represent the virtual-photon process resulting from the system without RWA.

In the interaction picture, the interaction Hamilton of atom-field coupling becomes
\begin{equation}\label{Effective Hamiltonian11}
H_I=g(\sigma_{+}a+a^{\dagger}\sigma_{-})+a^{\dagger}\sigma_{+}{\rm{e}}^{2{\rm{i}}\omega t}+a\sigma_{-}{\rm{e}}^{-2{\rm{i}}\omega t}.
\end{equation}
The evolution of the system characterized by the state vector $|\Psi(\pm\alpha,t)\rangle$ is given by the Sch\"odinger equation,
$$
{\rm{i}}\frac{\partial}{\partial t}|\Psi(\pm\alpha,t)\rangle=H_I|\Psi(\pm\alpha,t)\rangle,
$$
with its solution of the following form,
\begin{equation}\label{evolved state}
|\Psi(\pm\alpha,t)\rangle=\sum\limits_{n=0}^{\infty}[A_n(\pm\alpha,t)|e,n\rangle+B_n(\pm\alpha,t)|g,n\rangle].
\end{equation}
Substituting (\ref{evolved state}) into the Sch\"odinger equation, one has that
\begin{eqnarray}
{\rm{i}}\frac{\partial}{\partial t}A_n&=&g(B_{n+1}\sqrt{n+1}+B_{n-1}\sqrt{n}{\rm{e}}^{2{\rm{i}}\omega t}), \nonumber\\
{\rm{i}}\frac{\partial}{\partial t}B_n&=&g(A_{n-1}\sqrt{n}+A_{n+1}\sqrt{n+1}{\rm{e}}^{-2{\rm{i}}\omega t}).
\end{eqnarray}
The solution $|\Psi(\pm\alpha,t)\rangle$ depends on the coefficients $A_n(\pm\alpha,t)$ and $B_n(\pm\alpha,t)$ given by
\begin{eqnarray}\label{coefficient with rotation approximation2}
A_n(\pm\alpha,t)&=&C[A_n^0(\pm\alpha,t)+A_n'(\pm\alpha,t)],\nonumber\\
B_n(\pm\alpha,t)&=&C[B_n^0(\pm\alpha,t)+B_n'(\pm\alpha,t)],
\end{eqnarray}
where
\begin{eqnarray}\label{coefficient with rotation approximation1}
A_n^0(\pm\alpha,t)&=&-{\rm{i}}\sin(\sqrt{n+1}gt)F_{(n+1)}(\pm\alpha),\nonumber\\
B_n^0(\pm\alpha,t)&=&\cos(\sqrt{n}gt)F_n(\pm\alpha),\nonumber\\
A_n'(\pm\alpha,t)&=&g\frac{\sqrt{n}F_{n-1}(\pm\alpha)}{2}[\frac{1-{\rm{e}}^{{\rm{i}}K_1(n-1)t}}{K_1(n-1)}\nonumber\\
& &+\frac{1-{\rm{e}}^{{\rm{i}}K_2(n-1)t}}{K_2(n-1)}],\nonumber\\
B_n'(\pm\alpha,t)&=&g\frac{\sqrt{n+1}F_{n+2}(\pm\alpha)}{2}[\frac{1-{\rm{e}}^{-{\rm{i}}K_2(n+2)t}}{K_2(n+2)}\nonumber\\
& &-\frac{1-{\rm{e}}^{-{\rm{i}}K_1(n+2)t}}{K_1(n+2)}],
\end{eqnarray}
with $K_1(n)=2\omega+\sqrt{n}g$, $K_2(n)=2\omega-\sqrt{n}g$ and $C$ as a normalization factor.

For simplicity, we suppose that the terms $\frac{g\sqrt{n}}{K_1(n-1)}$, $\frac{g\sqrt{n}}{K_2(n-1)}$,  $\frac{g\sqrt{n+1}}{K_1(n+2)}$  and $\frac{g\sqrt{n+1}}{K_2(n+2)}$ are not large so that the perturbation theory can be applied.
$A_n^0$ and $B_n^0$ ($A_n'$ and $B_n'$) in (\ref{coefficient with rotation approximation2}) are zero ordered (first ordered) approximation corresponding to the solution of the JC model with (without) RWA in \cite{Peng1992PRA}.
$A_n'$ and $B_n'$ represent the influence of the virtual-photon processes on $A_n^0$ and $B_n^0$ \cite{Peng1992PRA}.
From Eqs. (\ref{coefficient with rotation approximation1}),
one can see that $A_n'$ and $B_n'$ remain infinitesimal only
when the radiation field is not intensive.
If all of the four terms all exist, that is the non-RWA case; otherwise, we will return to the case in \cite{Han2018NJP} with RWA if both $A_n'$ and $B_n'$ are neglected.

Since the discrimination of light states is dependent on the ancilla---the atoms, a von Neumann measurement on the atomic states can induce the successful discrimination
of the light states. Tracing over the variables of field, the atomic state is acquired as
\begin{eqnarray} \label{Atomic states}
\rho(\!\pm\alpha,t\!)\!&=&\!{\rm{Tr}}_F|\psi(\pm\alpha,t)\rangle\langle\psi(\pm\alpha,t)|\nonumber\\
\!&=&\!\sum\limits_{n\!=\!0}^{\infty}{\left(
\begin{array}{ccc}
  \!A_n\!A_n^*\! & \!A_nB_n^*\!\\
   \!A_n^*\!B_n\! & \!B_nB_n^*\!
\end{array}
\right)}.
\end{eqnarray}
The error probability for discriminating the two coherent states $|\alpha\rangle$ and $|-\alpha\rangle$ is lower bounded by
$$P_{\rm{err}}^{\min}=\min\limits_{t}\frac{1}{2}[1-2D(\alpha,t)],$$ with
\begin{equation}\label{trace disttance}
D(\alpha,t)\!\!=\!\!\max\limits_{\Pi}|{\rm{Tr}}\Pi[P_1\rho(\alpha,t)\!\!-\!\!P_2\rho(\!\!-\!\alpha,t)]|\!\!-\!\!\frac{1}{2}|P_1\!\!-\!\!P_2|.
\end{equation}
We call $D(\alpha,t)$ the trace distance which is closely related to the successful probability in discriminating the two coherent states.
The projective measurement operator for state discrimination can be written as $\Pi(r,\theta)=|\phi\rangle\langle\phi|$, where
$|\phi\rangle=\frac{1}{\sqrt{1+r^2}}(|g\rangle+{\rm{e}}^{\rm{i}\theta} r|e\rangle)$ with real parameters $r$ and $\theta$.
In order to minimize the failure probability (to maximize $D(\alpha,t)$), one should seek for the optimal measurement and the parameter $gt$. Through the contour-plot of $D$ versus the parameters $r$ and $\theta$, one can easily find that the optimal $D$ is obtained at $\theta=\pm\pi/2$ and $|r|=1$, which is the same result as in \cite{Han2018NJP} with RWA.

\begin{figure}
\includegraphics[width=8cm]{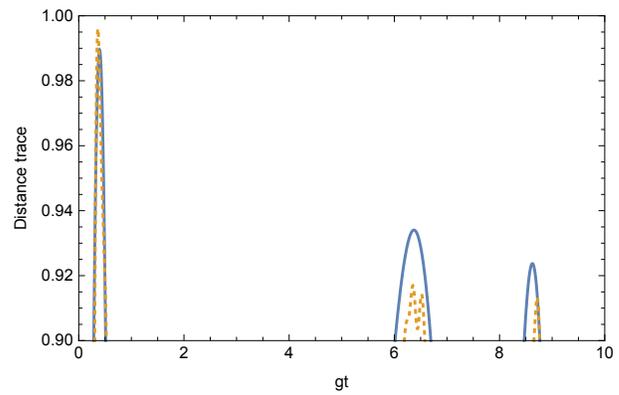} \\
 \caption{Trace distance $D[\rho(\alpha,t),\rho(-\alpha,t)]$ against $gt$ for $\alpha=2$, $\theta=\pi/2$ and $r=1$. The solid and dotted line correspond to the cases with and without RWA, respectively.} \label{fig1}
\end{figure}

Since it is difficult to have a large value of $gt$ experimentally, we chose the scope of $gt$ as $0\leq gt\leq10$ \cite{Han2018NJP}.
In Fig. \ref{fig1} we present the trace distance $D[(\alpha,t)]$ against $gt$.
Compared with the result with RWA in \cite{Han2018NJP}, the global optimal successful probability for identifying the two light states are increased from $0.9896$ (at $gt=0.3960$)
to $0.9960$ (at $gt=0.3636$). In addition, the result is enhanced obviously when the RWA is excluded for $\alpha=2$ rather than $\alpha=0.5,1$.

\begin{figure}
\includegraphics[width=8cm]{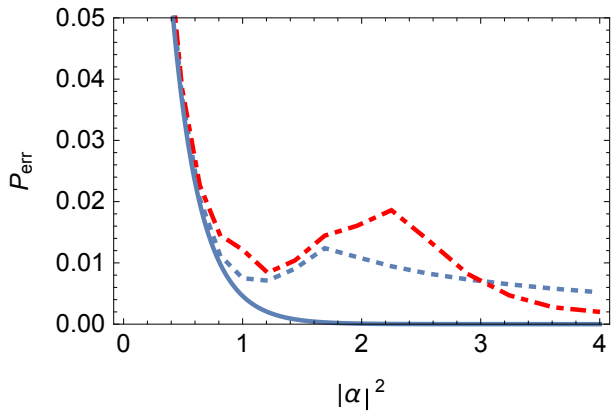} \\
\caption{The minimum error probability for ambiguous discrimination of coherent states $|\alpha\rangle$ and $|-\alpha\rangle$ versus the average photon number $|\alpha|^2$.
Solid line: the ideal Helstrom bound; dotted (dot-dashed) line: the minimum error probability for the discrimination of the two coherent states relying on JC model with (without) RWA, respectively.} \label{fig2}
\end{figure}

The minimum error probability $P_{\rm{err}}^{\min}$ of discriminating the two coherent states versus the average photon numbers $|\alpha|^2$ is shown in Fig. \ref{fig2}.
The result of the JC model without RWA is superior to the one with RWA for $|\alpha|^2>3$. Thus, the ideal Helstrom bound can be approached to a further step and one can acquire
higher precision of state discrimination via adjusting the average photon number of coherent states to a proper value.
Since the inner product of $|\alpha\rangle$ and $|-\alpha\rangle$ increases with the average photon number $|\alpha|^2$, one concludes that as the inner product of states increases,
the superiority of light state discrimination of JC model without RWA versus the one with RWA is enhanced, obviously.


The local maximum  $0.0196$ ($0.0118$) in the dot-dashed (dotted) line in Fig. \ref{fig2} can be acquired for JC model without (with) RWA at $gt=0.58$, $|\alpha|^2=2.35$ ($|\alpha|^2=1.65$, $gt=8.43$). These results, corresponding to the minimum optimal success probability, imply that the effect of the non-RWA term in Eq. (9) is impaired. One can draw an opposite conclusion for the model including RWA.

In contrast, with relative significance the local minimum of $P_{\rm{error}}$, $0.0084$ ($0.0069$) in the dot-dashed (dotted) line in Fig. \ref{fig2}, is acquired at $gt=8.41$, $|\alpha|^2=1.22$ ($gt=8.35$, $|\alpha|^2=1.15$) for the JC model without (with) RWA. From Fig. \ref{fig2}, it is seen that the difference between the local minimum of $P_{\rm{error}}$ for the JC model with and without RWA is much smaller than the one for local maxima.

Depending on detailed physical systems, the ideal bound (e.g. the Helstrom bound for ambiguous discrimination) may not be always saturated. Some information may still left in the post-measured states. The observer can perform a subsequent measurement to proceed the discrimination operation. This is called the non-destructive scheme. The larger the result of the first measurement deviates from the ideal bound, the more superiority this non-destructive scheme has. Since such state discrimination scheme with sequential measurement bases
on the failure of the first measurement, it completely differs from the sequential state discrimination (SSD) mentioned
in \cite{Fields2020arxiv} and \cite{Zhang2017arXiv,Bergou2003PRL} for ambiguous and unambiguous discriminations, respectively, where
the two measurements are preformed by different observers and unrelated to each other.

Since unambiguous state discrimination requires zero errors, there exist additional constraints on the POVM operators. Then, the unambiguous discrimination tends to be more possibly failed than the ambiguous one, see \emph{ Theorem 1} in the Supplemental Material. Then, we will present the implementation of unambiguous state discrimination with sequential measurement performed by Kennedy receiver as a special instance \cite{Han2018NJP}.

\section{Unambiguous state discrimination with sequential measurement.}\label{mixed local state discrimination}

In unambiguous discrimination of $N$ ($N\geq2$) nonorthogonal states, the projection measurements are applied to an extended space of at most $2N-1$ dimensions \cite{Roa2002PRA}.
Thus, the discrimination of coherent states $|\alpha\rangle$ and $|-\alpha\rangle$ can be implemented via JC model including three-level atoms, see the general scheme shown in the Supplemental Material. One may also tend to identify one of the states more successfully while ignore the another state which can be realized via a displacement operator. Such asymmetric
discrimination, called ``Kennedy-receiver scheme" \cite{Han2018NJP}, can be implemented physically via a two-level ancilla interacting with the light field via JC model without RWA. Below we present the scheme for nondestructive implementation of the Kennedy receiver via JC model without RWA that unambiguously discriminates the coherent states.

Before interacting with the auxiliary atoms, we let the light fields in states $|2\alpha\rangle$ and $|0\rangle$, obtained by applying the displacement operator $D(\alpha)$ to states $|\alpha\rangle$ and $|-\alpha\rangle$, respectively, interact with a sequence of atoms in ground state $|g\rangle$.
Suppose that the frequency and polarization of the coherent states matches with that of the atomic transition, so that the atom can be excited from the ground state $|g\rangle$ to the state $|e\rangle$. We can discriminate the states $|2\alpha\rangle$ and $|0\rangle$ unambiguously via a von Neumann measurement on the atomic states with respect to the basis \{$|g\rangle$, $|e\rangle$\}.


Since the atomic level transition is prohibited for the vacuum state $|0\rangle$ \cite{Han2018NJP}, the state $|0\rangle$ is bound to be neglected.
Then, the whole system evolves through the following unitary operator $U$ such that
\begin{eqnarray}
U|2\alpha\rangle|g\rangle&=&D(2\alpha, t)|\psi\rangle|e\rangle+E(2\alpha, t)|\phi\rangle|g\rangle,\nonumber\\
U|0\rangle|g\rangle&=&|0\rangle|g\rangle,
\end{eqnarray}
with the functions $D(2\alpha, t)$ and $E(2\alpha, t)$ satisfying
$$|D(2\alpha, t)|^2=\sum\limits_{n=0}^{\infty}|A_n(2\alpha,t)|^2,$$
$$|E(2\alpha, t)|^2=\sum\limits_{n=0}^{\infty}|B_n(2\alpha,t)|^2,$$
where the functions $A_n(2\alpha, t)$ and $B_n(2\alpha, t)$ are of the same form as in Eq. (\ref{evolved state}) with the parameters $\pm\alpha$ replaced by $2\alpha$.

Then, we can identify the state $|2\alpha\rangle$ through a von Neumann measurement $|e\rangle\langle e|$ on the auxiliary system.
If the atom is found to be in the exited state (atomic level transition is successful), the discrimination is successful. The post-measured light states $|\psi\rangle$ and $|\phi\rangle$ correspond to $|e\rangle$ and $|g\rangle$, respectively.



If the atom is found to be in the ground atomic state $|g\rangle$, we perform a sequential measurement on the ancilla to identify the post-measured state $|\phi\rangle$.
Suppose that the optimal success probability of the first measurement, denoted by excitation probability of the atom, is $P_A=\max\limits_{t}|D(2\alpha,t)|^2$ that is acquired at $t=t_0$. By straightforward calculation we have
$$|\phi\rangle=\frac{1}{\sqrt{\sum\limits_{n=0}^{\infty}|B_n(2\alpha,t_0)|^2}}\sum\limits_{n=0}^{\infty}B_n(2\alpha,t_0)|n\rangle.$$

The state of the atom-field system is given by
$$
|\Psi(t')\rangle=\sum\limits_{n=0}^{\infty}[\tilde{A}_n(2\alpha,t')|e,n-1\rangle+\tilde{B}_n(2\alpha,t')|g,n\rangle],
$$
where
\begin{eqnarray}
\tilde{A}_n(2\alpha,t')&=&C'[\tilde{A}_n^0(2\alpha,t')+\tilde{A}_n'(2\alpha,t')],\nonumber\\
\tilde{B}_n(2\alpha,t')&=&C'[\tilde{B}_n^0(2\alpha,t')+\tilde{B}_n'(2\alpha,t')],
\end{eqnarray}
with
\begin{eqnarray}\label{coefficient with rotation approximation}
\tilde{A}_n^0(2\alpha,t')&=&-{\rm{i}}\sin(\sqrt{n+1}gt')B_{(n+1)}(2\alpha,t_0),\nonumber\\
\tilde{B}_n^0(2\alpha,t')&=&\cos(\sqrt{n}gt')B_n(2\alpha,t_0),
\end{eqnarray}
and
\begin{eqnarray}\label{coefficient without rotation approximation}
\tilde{A}_n'(2\alpha,t')&=&\frac{\sqrt{n}gB_{n-1}(2\alpha,t_0)}{2}[\frac{1-{\rm{e}}^{{\rm{i}}K_1(n-1)t'}}{K_1(n-1)}\nonumber\\
& &+\frac{1-{\rm{e}}^{{\rm{i}}K_2(n-1)t'}}{K_2(n-1)}],\nonumber\\
\tilde{B}_n'(2\alpha,t')&=&\frac{\sqrt{n+1}gB_{n+2}(2\alpha,t_0)}{2}[\frac{1-{\rm{e}}^{-{\rm{i}}K_2(n+2)t'}}{K_2(n+2)}\nonumber\\
& &-\frac{1-{\rm{e}}^{-{\rm{i}}K_1(n+2)t'}}{K_1(n+2)}],
\end{eqnarray}
$K_1(n)=2\omega+\sqrt{n}g$, $K_2(n)=2\omega-\sqrt{n}g$ and $C'$ is a normalization factor.

The subsequent measurement identifies $|\phi\rangle$ with an optimal successful probability $P_B=\max\limits_{t'}\sum\limits_{n=0}^{\infty}|\tilde{A}_n(2\alpha,t')|^2$.
Then, see \emph{Theorem 2} in the Supplemental Material,  the total failure probability of these two sequential measurements is given by
$$Q_{\rm{SM}}=1-P_1[P_A+(1-P_A)P_B].$$

\begin{figure}
\includegraphics[width=8cm]{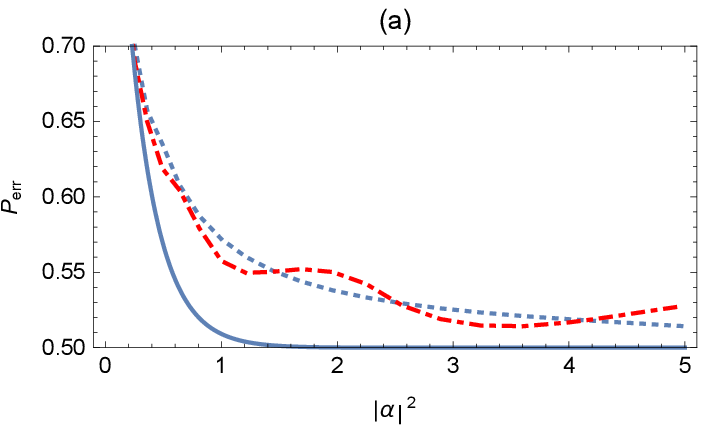} \\
\includegraphics[width=8cm]{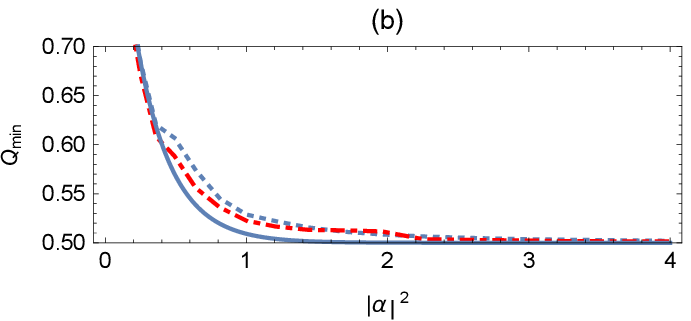} \\
 \caption{The minimum failure probability $Q_{\min}$ of unambiguous discrimination of coherent states $|2\alpha\rangle$ and $|0\rangle$ against the average photon number $|\alpha|^2$ with only first measurement in (a) and
 two sequential measurements in (b). Solid line: the ideal Kennedy bound; dotted (dot-dashed) line: the error probability for JC-model discrimination of
 the two coherent states with (without) RWA, respectively.} \label{fig3}
\end{figure}

The minimum failure probability with only one measurement and two sequential measurements are shown
as functions of the average photon number $|\alpha|^2$ in Fig. \ref{fig3} (a) and (b), respectively.
The ideal Kennedy bound,
\begin{equation}\label{ideal Kennedy bound}
Q_{\min}=P_1+P_2{\rm{e}}^{-4|\alpha|^2},
\end{equation}
which is calculated in the Supplemental Material, is also shown in Fig. \ref{fig3}.

It is seen that for the unambiguous state discrimination implemented via the JC model without RWA,
the result is superior to the one with RWA except for the cases where the average photon numbers $|\alpha|^2$ approach $2$. Compared with the results of optimal ambiguous state
discrimination, the superiority of the non-RWA JC model
versus the one with RWA is more obvious, see Figs. \ref{fig2} and \ref{fig3} (a).
In Fig. \ref{fig3} (a), there exist two local minima $0.5507$ (at $|\alpha|^2=1.15$, $gt=0.70$) and 0.5142 (at $|\alpha|^2=3.56$, $gt=0.38$) in the dot-dashed line. The local maximum also implies the weakening of the non-RWA term under the framework of unambiguous state discrimination with only one measurement. In contrary, with respect to the local minimum $0.5142$, the effect of non-RWA term is strengthened, in coinciding with the global minimum of $Q_{\rm{\min}}$. It is also shown that the minimum failure probability  $Q_{\rm{\min}}$ for the non-RWA JC model will increase asymptotically with $|\alpha|^2$ when the average photon number is larger than the value corresponding to the global minimum. The failure probability will surpass the one with RWA again as $|\alpha|^2>4.07$.

As the number of measurements increases, from Fig. \ref{fig3} (b) we have that the superiority of the results for the JC model without RWA is further enhanced.
Namely, QEOF in the JC model without RWA may enhance the discrimination of quantum states, especially for the protocol with sequential measurements.

\begin{figure}
\includegraphics[width=8cm]{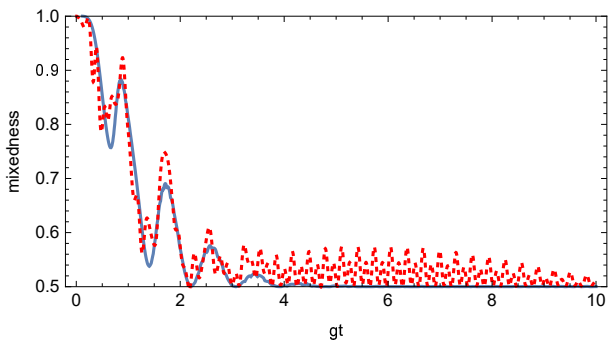} \\
\caption{The purity $F(\rho)$ of the auxiliary state against $gt$ for $|\alpha|^2=3.56$ (optimal global minimum of $Q_{\rm{min}}$). Solid (dotted) line corresponds to the
result with (without) RWA.} \label{fig4}
\end{figure}

The protocol based on sequential measurement includes an initial pure ancilla state which incurs additional resource overhead. We consider the cost of resource via the purity of the auxiliary states $F(\rho)={\rm{Tr}}({\rho}^2)$, see Fig. \ref{fig4}. It is shown that the result of non-RWA JC model has a larger purity of the ancilla than the one with RWA. Consequently, one can conclude that unambiguous state discrimination implemented via the non-RWA JC model is beneficial to saving resource cost.


\section{Conclusion}\label{Summ}

In summary, we have investigated the physical implementations of both the ambiguous and unambiguous state discriminations via JC interaction
between the light field and atoms without RWA. It has been shown that for ambiguous state discriminations, only for larger phonon numbers the results for the JC model without RWA is superior to the one with RWA. For unambiguous state discrimination, we adopted the so-called Kennedy receiver protocol including sequential measurements.
Compared with the results of ambiguous discrimination, the superiority of unambiguous state discrimination
without RWA is enhanced compared with the one with RWA. This superiority becomes more obvious as the number of sequential measurements increases.
These results may be attributed to the fact that the non-RWA term in Eq. (\ref{Effective Hamiltonian1}) bring about the virtual-photon processes including QEOF in the JC model without RWA which play a critical role in enhancing the discrimination of quantum states, especially for the protocol with sequential measurements.
We also find that unambiguous state discrimination implemented via the non-RWA JC model is beneficial to saving resource cost. Since the non-RWA JC model always corresponds to a strong-coupling regime which is closer to a real practical physical system with respect to the actual experimental setups \cite{Wallraff2004Nature,Simmonds2004PRL,Chiorescu2004Nature,Johansson2006PRL,Yu2003Science,Forn2010PRL,Liu2009Eru}, it is desirable that our results can be verified experimentally.

\emph{Acknowledgments.} This work was funded by the NSF of China (Grant Nos. 11675119, 12075159, 11575125, 12171044), Shanxi Education Department Fund (2020L0543), Beijing Natural Science Foundation (Z190005),
Academy for Multidisciplinary Studies, Capital Normal University, the Academician Innovation Platform of Hainan Province,
and the Shenzhen Institute for Quantum Science and Engineering, Southern University of Science and Technology (No. SIQSE202001).

\appendix
\section{General Schemes for Unambiguous State Discrimination} \label{mixed non-orthogonal state}
(1) \emph{Unambiguous state discrimination with one measurement.}

In order to discriminate the state $|\psi_1\rangle$ from $|\psi_2\rangle$ ($0<\langle\psi_1|\psi_2\rangle=s<1$) unambiguously,
we couple the system with a three-level ancilla state $|0_b\rangle$ and perform a joint unitary transformation $U_b$ \cite{Zhang2017arXiv} such that
\begin{equation}\label{Ub1}
U_b|\psi_i\rangle|0_b\rangle=\sqrt{q_i^b}|\chi_i\rangle|0\rangle_b+\sqrt{1-q_i^b}|\phi_i\rangle|i\rangle_b,
\end{equation}
with $0<\langle\chi_1|\chi_2\rangle=t\leq1$, and perform a von Neumann measurement with respect to the basis \{$|0\rangle$, $|1\rangle$, $|2\rangle$\}
on the auxiliary qutrit.

The discrimination is successful if the ancilla collapses to the states $|1\rangle_b$ or $|2\rangle_b$, and failed if the outcome state is
$|0\rangle_b$. Hence the average failure probability is given by
\begin{equation}\label{Pb1}
Q^{(1)}=P_1q_1^b+P_2q_2^b,
\end{equation}
with the constraint $q_1q_2=s^2/t^2$ as $U_b$ is unitary. The lower bound of the average failure probability is then given by
\begin{equation}\label{Pb22}
Q^{(1)}_{\rm{min}}=2\sqrt{P_1P_2}s,
\end{equation}
which is saturated at $t=1$ (optimal measurement) and $q_1^b=\sqrt{P_2/P_1}s$
under the condition that $0<s<\sqrt{P_1/P_2}$.  For $\sqrt{P_1/P_2}\leq s<1$,
this lower bound reduces to $P_1+P_2s^2$ achieved at $q_b=1$, $t=1$ with one state $|\Psi_1\rangle$ bound to be neglected.


In the physical implementation of unambiguous state discrimination  performed
by Kennedy receiver mentioned in the main text, only two-level ancilla is allowed.
Thus, one state has to be ignored which implies that $q_1^b=1$. Hence, the minimum failure probability can be acquired via the following relations:
\begin{equation}
{\rm{minimize:}}\
Q^{(2)}=P_1+P_2q_2^b,
\end{equation}
\begin{eqnarray}\label{conditional}
{\rm{subject\ to:}}\
q_2^b\in[s^2/t^2,1].
\end{eqnarray}
Thus, the lower bound $Q^{(2)}_{\rm{min}}=P_1+P_2s^2$ is saturated at $t=1$ and $q_2^b=s^2$, which is of the same form as the one with three-level ancilla
for $\sqrt{P_1/P_2}\leq s<1$. Then, by setting $|\psi_1\rangle=|\alpha\rangle$ and $|\psi_2\rangle=|-\alpha\rangle$, the relation $Q_{\min}=P_1+P_2{\rm{e}}^{-4|\alpha|^2}$ can be acquired easily. Comparing the results of ambiguous and unambiguous state discrimination, we have the following theorem.

\emph{Theorem 1}.\ The minimum failure probability of unambiguous state discrimination is superior to the one of ambiguous state discrimination.

\emph{Proof.} The difference of minimum failure probability between ambiguous and unambiguous discrimination satisfies
\begin{eqnarray}
\Delta Q&=&Q^{(1)}_{\min}-P_{\rm{err}}\nonumber\\
&=&2\sqrt{P_1P_2}s-\frac{1}{2}(1-\sqrt{1-4P_1P_2s^2})\nonumber\\
&=&Q^{(1)}_{\min}+\frac{1}{2}\sqrt{1-Q^{(1)2}_{\min}}-\frac{1}{2}\nonumber\\
&\geq&\sqrt{\frac{1}{4}+\frac{3}{4}Q^{(1)2}_{\min}}-\frac{1}{2}
\geq 0.
\end{eqnarray}
Therefore, we have
$$Q_{\min}^{(2)}=P_1+P_2s^2\geq2\sqrt{P_1P_2}s=Q_{\min}^{(1)}\geq P_{\rm{err}}.$$
$\Box$

(2) \emph{Unambiguous state discrimination with sequential measurement.}

Since the ideal bound for state discrimination can not always be saturated in detailed physical implementation, the state discrimination operation in
Eq. (\ref{Ub1}) is not always optimal (e.g. the general case $\langle\phi_1|\phi_2\rangle=t<1$).
If the discrimination succeeds via the first measurement, this procedure ends. Otherwise, we couple the system with another three-level ancilla state $|0_c\rangle$
and continue to discriminate the
post-measured state via another unitary operation $U_c$,
\begin{equation}\label{Ub11}
U_c|\chi_i\rangle|0_c\rangle=\sqrt{q_i^c}|\chi'\rangle|0\rangle_c+\sqrt{1-q_i^c}|\phi'_i\rangle|i\rangle_c,
\end{equation}
and a subsequent von Neumann measurement on the auxiliary qutrit.
Then, the success probability with respect to the first (second) measurement $P_A$ ($P_B$) is given by
\begin{eqnarray}\label{success probability 1}
P_A&=&1-Q^{(1)}=P_1(1-q_1^b)+P_2(1-q_2^b), \nonumber\\
P_B&=&P_1^0(1-q_1^c)+P_2^0(1-q_2^c),
\end{eqnarray}
where $P_i^0=\frac{P_iq_i^b}{P_1q_1^b+P_2q_2^b}$ ($i=1,2$) are the conditional prior probability of the post-measured states
corresponding to the failure result of the first measurement. Hence, according to Eq. (\ref{success probability 1}), we have the following theorem.

\emph{Theorem 2}.\ The total failure probability $Q^{\rm{SM}}$ of the non destructive scheme with sequential measurement is given by
\begin{eqnarray}\label{success probability 2}
Q^{\rm{SM}}&=&P_1q_1^bq_1^c+q_2^bq_2^c\nonumber\\
&=&1-P_1[P_A+(1-P_A)P_B].
\end{eqnarray}

In Ref. \cite{JHZhang2020PRA}, the two measurements are performed by different observers. Since the latter measurement occurs on the premise that the first one fails, the
classical communications between the two observers are needed. However, in our scheme realized via non destructive implementation mentioned in the main text, the classical communications are not necessary. Then, according to the constraints $0<q_1^c,q_2^c\leq1$, $q_1^cq_2^c\leq t^2$ and the relation (\ref{success probability 2}), we have the following corollary.

\emph{Corollary}. If the first measurement is optimal ($t=1$), we have $Q^{\rm{SM}}=Q^{(1)}$. Otherwise, if $t<1$ we have $Q^{\rm{SM}}<Q^{(1)}$.




\begin{thebibliography}{99}\footnotesize
\itemsep=-1pt plus.2pt minus.2pt

\bibitem{Enk2002PRA}
van Enk S J 2002 Unambiguous state discrimination of coherent states with linear
optics: Application to quantum cryptography {\it Phys. Rev. A.} {\bf 66} 042313
\bibitem{Helstrom1976}
Helstrom C  W 1976 {\it Quantum detection and estimation theory} (New York, NY: Academic Press)
\bibitem{Wittmann2008}
Wittmann C,  Takeoka M, Cassemiro K  N,  Sasaki M,  Leuchs  G and  Andersen U  L  2008  Demonstration of Near-Optimal Discrimination of Optical Coherent
States {\it Phys. Rev. Lett.} {\bf 101}
210501
\bibitem{Tsujino2011}
Tsujino K,  Fukuda D,  Fujii G,  Inoue S,  Fujiwara M,
Takeoka M and Sasaki M 2011  Quantum Receiver beyond the Standard Quantum Limit of Coherent
Optical Communication {\it Phys. Rev. Lett.}  {\bf 106} 250503
\bibitem{Assalini2011}
Assalini A,  Pozza N  D and  Pierobon G 2011 Revisiting the Dolinar receiver
through multiple-copy state discrimination theory {\it Phys. Rev. A. } {\bf 84}  022342
\bibitem{Xiong2018JPA}
Xiong C and  Wu J 2018 Geometric coherence and quantum state discimination {\it J. Phys. A Math. Theor.}  {\bf 51}  414005
\bibitem{Xiong2019PRA}
Xiong C,  Kumar  A,  Huang. M,  Das. S, Sen. U and Wu. J  2019  Partial coherence and quantum correlation with fidelity and affinity
distances {\it Phys. Rev. A.}   {\bf 99}   032305
\bibitem{Fields2020arxiv}
Fields Dov,  Varga \'{A}rp\'{a}d and  Bergou J\'{a}nos  A 2020 Sequential measurements on qubits by multiple observers: Joint best guess strategy {\it arXiv:2005.11656v1} {\bf 24}
\bibitem{Ivanovic1987PLA}
Ivanovic I D 1987  How to differentiate between non-orthogonal states {\it Phys. Lett. A.}  {\bf 123}  257
\bibitem{Peres1988PLA}
Peres  A 1988 How to differentiate between non-orthogonal states {\it  Phys. Lett. A.}  {\bf 128}  19
\bibitem{Dies1988PLA}
Dieks D 1988  Overlap and distinguishability of quantum states {\it  Phys. Lett. A.}  {\bf 126}  303
\bibitem{Bennett1992PRL}
Bennett C H 1992  Quantum Cryptography Using Any Two Nonorthogonal States {\it Phys. Rev. Lett.}  {\bf 68}  3121
\bibitem{Bergou2003PRL}
Bergou J A,  Herzog  U and Hillery  M 2003  Quantum Filtering and Discrimination between Sets of Boolean Functions {\it Phys. Rev. Lett.}
{\bf 90}  257901
\bibitem{Pang2009PRA}
Pang  S and Wu  S 2009 Optimum unambiguous discrimination of linearly independent
pure states {\it Phys. Rev. A}  {\bf 80}  052320
\bibitem{Bergou2013PRL}
Bergou  J, Feldman  E and Hillery  M 2013 Extracting Information from a Qubit by
Multiple Observers: Toward a Theory of Sequential State Discrimination {\it Phys. Rev. Lett.}
{\bf 111}  100501
\bibitem{Namkung2017PRA}
Namkung  M and Kwon  Y 2017 Optimal sequential state discrimination between two
mixed quantum states {\it Phys. Rev. A.} {\bf 96} 022318
\bibitem{Pang2013PRA}
Pang  C-Q, Zhang  F-L, Xu  L-F, Liang M-L and Chen  J-L 2013 Sequential state discrimination and requirement of quantum dissonance {\it Phys. Rev. A}  {\bf 88} 052331
\bibitem{Namkung2018SR}
Namkung  M and Kwon  Y 2018 Analysis of Optimal Sequential State Discrimination for Linearly Independent Pure Quantum States {\it Sci. Rep.}  {\bf 8}  6515
\bibitem{Zhang2017arXiv}
Zhang  J-H, Zhang F-L and Liang  M-L 2018 Sequential state discrimination with
quantum correlation {\it Quantum. Inf. Process.} {\bf 17}  260
\bibitem{JHZhang2020PRA}
Zhang J-H, Zhang F-L, Wang Z-X, Lai L-M and Fei S-M 2020 Discrimination bipartite mixed states by local operations {\it Phys. Rev. A.} {\bf 101} 032316
\bibitem{JHZhang2020Entropy}
Zhang J-H,   Zhang F-L,   Wang  Z-X, Yang H and Fei S-M 2022 Unambiguous state discrimination with intrinsic coherence {\it Entropy.} {\bf 24}  18
\bibitem{Silberhorn2002PRL}
Silberhorn  C, Ralph T C, L¨¹tkenhaus N and Leuchs G 2002 Continuous Variable Quantum Cryptography: Beating the 3 dB Loss Limit {\it Phys. Rev. Lett.} {\bf 89} 167901
\bibitem{Lorenz2004APB}
 Lorenz S, Korolkova N and Leuchs  G 2004 Continuous-variable quantum key distribution using polarization encoding and post selection  {\it Appl. Phys. B.} {\bf 79}  273
\bibitem{Lance2005PRL}
Lance A M, Symul T, Sharma V, Weedbrook C, Ralph T C and Lam P K 2005 No-Switching Quantum Key Distribution Using Broadband Modulated Coherent Light {\it Phys. Rev. Lett.} {\bf 95} 180503
\bibitem{Takeoka2005PRA}
Takeoka M, Sasaki M, van Loock P and L¨¹tkenhaus N 2005 Implementation of projective measurements with linear optics and continuous photon counting {\it Phys. Rev. A}  {\bf 71}  022318
\bibitem{Takeoka2008PRA}
 Takeoka  M and Sasaki  M 2008 Discrimination of the binary coherent signal: Gaussian-operation limit and simple non-Gaussian near-optimal receivers {\it Phys. Rev. A} {\bf 78} 022320
\bibitem{Bergou2010JMO}
Bergou. J. A, 2010 Discrimination of quantum states {\it J. Mod. Opt.} {\bf 57}  160
\bibitem{Wittmann2010PRL}
Wittmann C, Andersen U L, Takeoka M, Sych D and Leuchs G 2010 Demonstration of Coherent-State Discrimination Using a Displacement-Controlled Photon-Number-Resolving Detector {\it Phys. Rev. Lett.}  {\bf 104} 100505
\bibitem{Wittmann2010PRA}
Wittmann C and Andersen U L 2010 Discrimination of binary coherent states using a homodyne detector and a photon number resolving detector {\it Phys. Rev. A.}  {\bf 81} 062338
\bibitem{Weedbrook2012RMP}
Weedbrook  C, Pirandola  S, Garcia-Patr¨®n  R, Cerf  N  J, Ralph  T  C, Shapiro  J  H and Lloyd  S
2012 Gaussian Quantum Information {\it Rev. Mod. Phys.}  {\bf 84}  621
\bibitem{Becerra2013Nap}
Becerra F E, Fan J, Baumgartner G, Goldhar J, Kosloski J T and Migdall A 2013 Experimental demonstration of a receiver beating the standard quantum limit for multiple nonorthogonal state discrimination {\it Nat. Photon.}  {\bf 7}  147
\bibitem{Becerra2014Nap}
Becerra F E, Fan J and Migdall A 2014 Photon number resolution enables quantum receiver for realistic coherent optical communications {\it Nat. Photon.} {\bf 9} 48
\bibitem{Sych2016PRA}
Sych D and Leuchs G 2016 Practical Receiver for Optimal Discrimination of Binary Coherent Signals {\it Phys. Rev. Lett.}  {\bf 117} 200501
\bibitem{Wittmann2010JMO}
Wittmann  C, Andersen  U  L and Leuchs  G  2010 Discrimination of optical coherent states using a photon number resolving detector {\it J. Mod. Opt.} {\bf 57}  213
\bibitem{Tsujino2011PRL}
 Tsujino K, Fukuda D, Fujii G, Inoue S, Fujiwara M, Takeoka M and Sasaki M 2011 Quantum Receiver beyond the Standard Quantum Limit of Coherent Optical Communication
 {\it Phys. Rev. Lett.} {\bf 106}  250503
\bibitem{M¨¹ller2015NJP}
M¨¹ller C R and Marquardt Ch 2015 A robust quantum receiver for phase shift keyed signals  {\it New J. Phys.}  {\bf 17}  032003
\bibitem{Han2018NJP}
Han  R,  Bergou  J  A and Leuchs  G 2018 Near optimal discrimination of binary coherent
signals via atom-light interaction {\it New J. Phys.} {\bf 20}  043005
\bibitem{Peng1992PRA}
Peng J and Li G 1992 Phase fluctuations in the Jaynes-Cummings model with and without the rotating wave approaximation {\it Phys. Rev. A.} {\bf 45}  3289
\bibitem{Peng1993PRA}
Peng J and Li G 1993 Influnce of the virtual-photon processes on the squeesing of light in the two photon Jaynes-Cummings model {\it Phys. Rev. A.} {\bf 47}  3167
\bibitem{Wallraff2004Nature}
Wallraff  A  et al 2004 Strong coupling of a single photon to a superconducting qubit using circuit quantum electrodynamics {\it Nature (London).}  {\bf 431} 162
\bibitem{Simmonds2004PRL}
Simmonds R W, Lang K M, Hite D A, Nam S, Pappas D P and
Martinis J M 2004 Decoherence in Josephson Phase Qubits from Junction Resonators {\it Phys. Rev. Lett.} {\bf 93} 077003
\bibitem{Chiorescu2004Nature}
Chiorescu I et al 2004 Coherent dynamics of a flux qubit coupled to a harmonic oscillator {\it Nature (London).} {\bf 431} 159
\bibitem{Johansson2006PRL}
Johansson J, Saito S,  Meno T, Nakano H,  Ueda M,
Semba K and Takayanagi H 2006 Vacuum Rabi Oscillations in a Macroscopic Superconducting Qubit LC Oscillator System {\it Phys. Rev. Lett.} {\bf 96} 127006
\bibitem{Forn2010PRL}
Forn-D¨ªaz P,  Lisenfeld J,  Marcos D, Garc¨ªa-Ripoll J  J, Solano E, Harmans C  J  P  M and Mooij J E 2010 Observation of the Bloch-Siegert Shift in a Qubit-Oscillator System in the Ultrastrong Coupling Regime {\it Phys. Rev. Lett.} {\bf 105}  237001
\bibitem{Yu2003Science}
Yu Y \emph{et al} 2002 Coherent Temporal Oscillations of Macroscopic Quantum States in a Josephson Junction  {\it Science.} {\bf 296} 889
\bibitem{Liu2009Eru}
Liu T, Wang K L and  Feng M 2009 The generalized analytical approximation to the solution of the single-mode spin-boson model without rotating-wave approximation {\it Europhys. Lett.} {\bf 86} 54003
\bibitem{Roa2002PRA}
Roa L, Retamal J and Saavedra C 2002 Quantum-state discrimination {\it Phys. Rev. A.} {\bf 66}  012103
\end{thebibliography}
\end{document}